\documentclass[journal,letterpaper]{IEEEtran_nssmic}
\usepackage{graphicx}  
\usepackage{amsmath}   

\begin{document}

\title{New Results on DEPFET Pixel Detectors for Radiation Imaging and High Energy Particle
Detection}

\author{\authorblockN{
N. Wermes\authorrefmark{1}, L. Andricek\authorrefmark{3}, P.
Fischer\authorrefmark{2},
K. Heinzinger\authorrefmark{4} S. Herrmann\authorrefmark{3}, M.
Karagounis\authorrefmark{1}, R. Kohrs\authorrefmark{1}, \\ H.
Kr\"uger\authorrefmark{1}, G. Lutz\authorrefmark{3}, P.
Lechner\authorrefmark{4}, I. Peric\authorrefmark{1}, M.
Porro\authorrefmark{3}, R.H. Richter\authorrefmark{3}, G.
Schaller\authorrefmark{3}, M.Schnecke-Radau\authorrefmark{3}, \\
F. Schopper\authorrefmark{3}, H. Soltau\authorrefmark{4}, L.
Str\"uder\authorrefmark{3}, M. Trimpl\authorrefmark{1}, J.
Ulrici\authorrefmark{1}, J. Treis\authorrefmark{3}}\\ \hfill \\
\authorblockA{\authorrefmark{1}Physikalisches Institut der Universit\"at Bonn, D-53115 Bonn,
Germany}\\
\authorblockA{\authorrefmark{2}Universit\"at Mannheim, D 7,
3-4, D-68159 Mannheim, Germany}\\
\authorblockA{\authorrefmark{3}MPI Hableiterlabor, Otto-Hahn-Ring
6, D-81739 M\"unchen, Germany}\\
\authorblockA{\authorrefmark{4}PN Sensor GmbH, R\"o merstr. 28, D-80803 Muenchen, Germany} \\
\thanks{Corresponding author: N. Wermes, email: wermes@physik.uni-bonn.de}
\thanks{Work supported by the German Ministerium f{\"u}r Bildung,
              Wissenschaft, Forschung und Technologie (BMBF) under contract
              no.~05 HA8PD1, by the
              Ministerium f{\"u}r Wissenschaft und Forschung des Landes
              Nordrhein--Westfalen under contract no.~IV A5-106 011 98, and
              by the Deutsche Forschungsgemeinschaft DFG}}


\maketitle

\begin{abstract}
DEPFET pixel detectors are unique devices in terms of energy and
spatial resolution because very low noise (ENC = 2.2e at room
temperature) operation can be obtained by implementing the
amplifying transistor in the pixel cell itself. Full DEPFET pixel
matrices have been built and operated for autoradiographical
imaging with imaging resolutions of 4.3$\pm$0.8 $\mu$m at 22 keV.
For applications in low energy X-ray astronomy the high energy
resolution of DEPFET detectors is attractive. For particle
physics, DEPFET pixels are interesting as low material detectors
with high spatial resolution. For a Linear Collider detector the
readout must be very fast. New readout chips have been designed
and produced for the development of a DEPFET module for a pixel
detector at the proposed TESLA collider (520x4000 pixels) with 50
MHz line rate and 25 kHz frame rate. The circuitry contains
current memory cells and current hit scanners for fast pedestal
subtraction and sparsified readout. The imaging performance of
DEPFET devices as well as present achievements towards a DEPFET
vertex detector for a Linear Collider are presented.
\end{abstract}


%

\section{Introduction}
For a Linear Collider Detector \cite{teslatdr} efficient and good
flavor identification and separation is required for the accurate
determination of Higgs branching ratios and other physics
processes beyond the Standard Model. However, due to the very
prominent beamstrahlung near the interaction point, the background
conditions and the time structure of the accelerator
are fierce leading to detector occupancies of 80 hits / mm$^2$ /
bunch train ($\sim$ 1ms) for a typical pixel detector situated at
a radius of 15 mm away from the beam line. DEPFET pixel detectors
have shown to be capable of simultaneously providing excellent
energy and spatial resolution which can be exploited for various
applications: imaging in biomedical autoradiography
\cite{neeser00,ulrici00,ulrici03}, imaging of low energy X-rays
from astronomical sources
\cite{holl00,klein00,strueder00,strueder03} and particle detection
at a future Linear Collider \cite{PRC03,richter02,trimpl02}. In
this paper we present the achievements obtained to date in imaging
and spectroscopy with DEPFET pixel devices and describe the
developments for a micro vertex detector based on DEPFET pixels
for a Linear Collider. The DEPFET developments for the XEUS
project are presented by L. Str\"uder \cite{strueder03} at this
conference.

\section{The DEPFET principle and operation}
\begin{figure}[h]
\begin{center}
\includegraphics[width=0.22\textwidth]{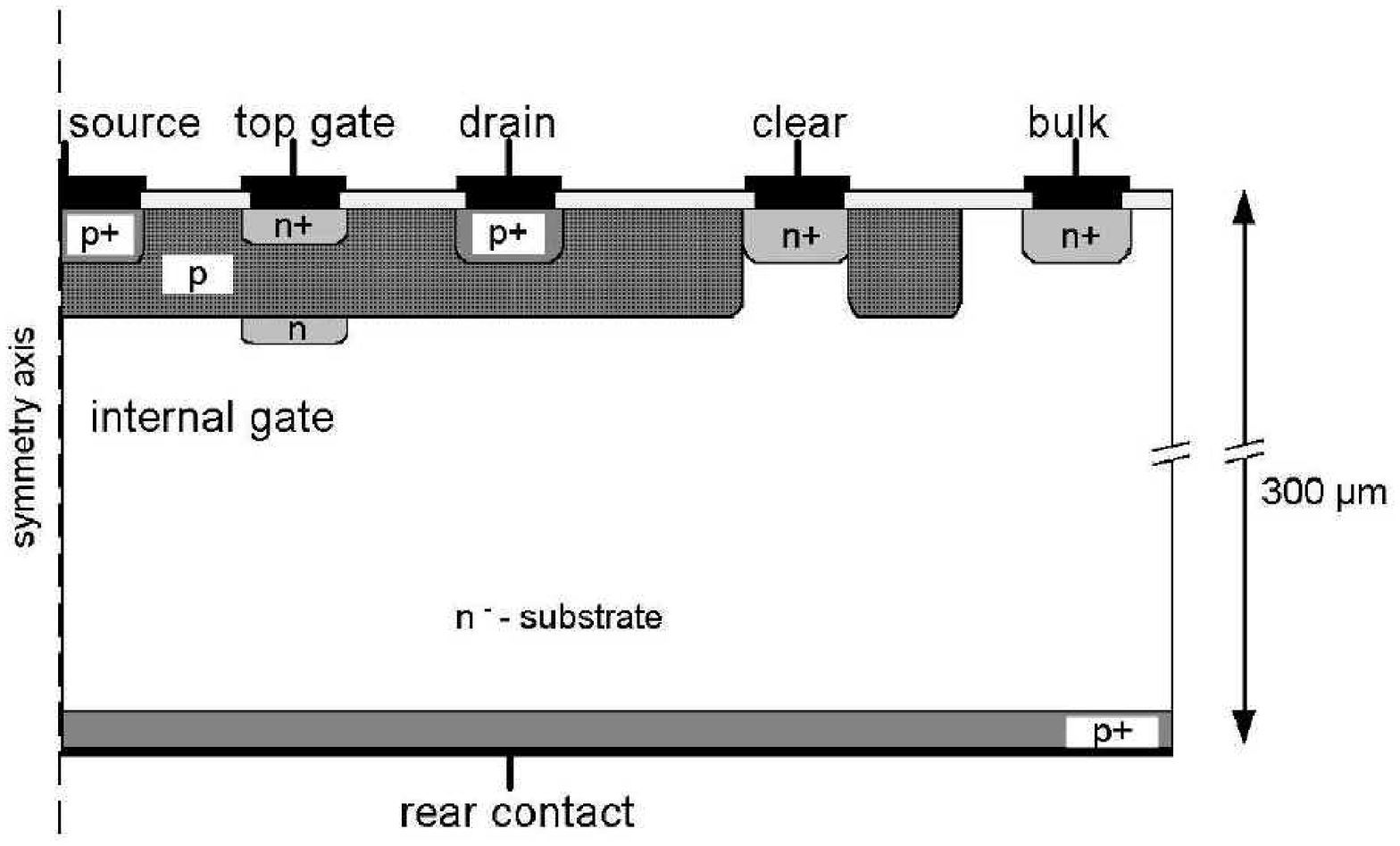}
\includegraphics[width=0.2\textwidth]{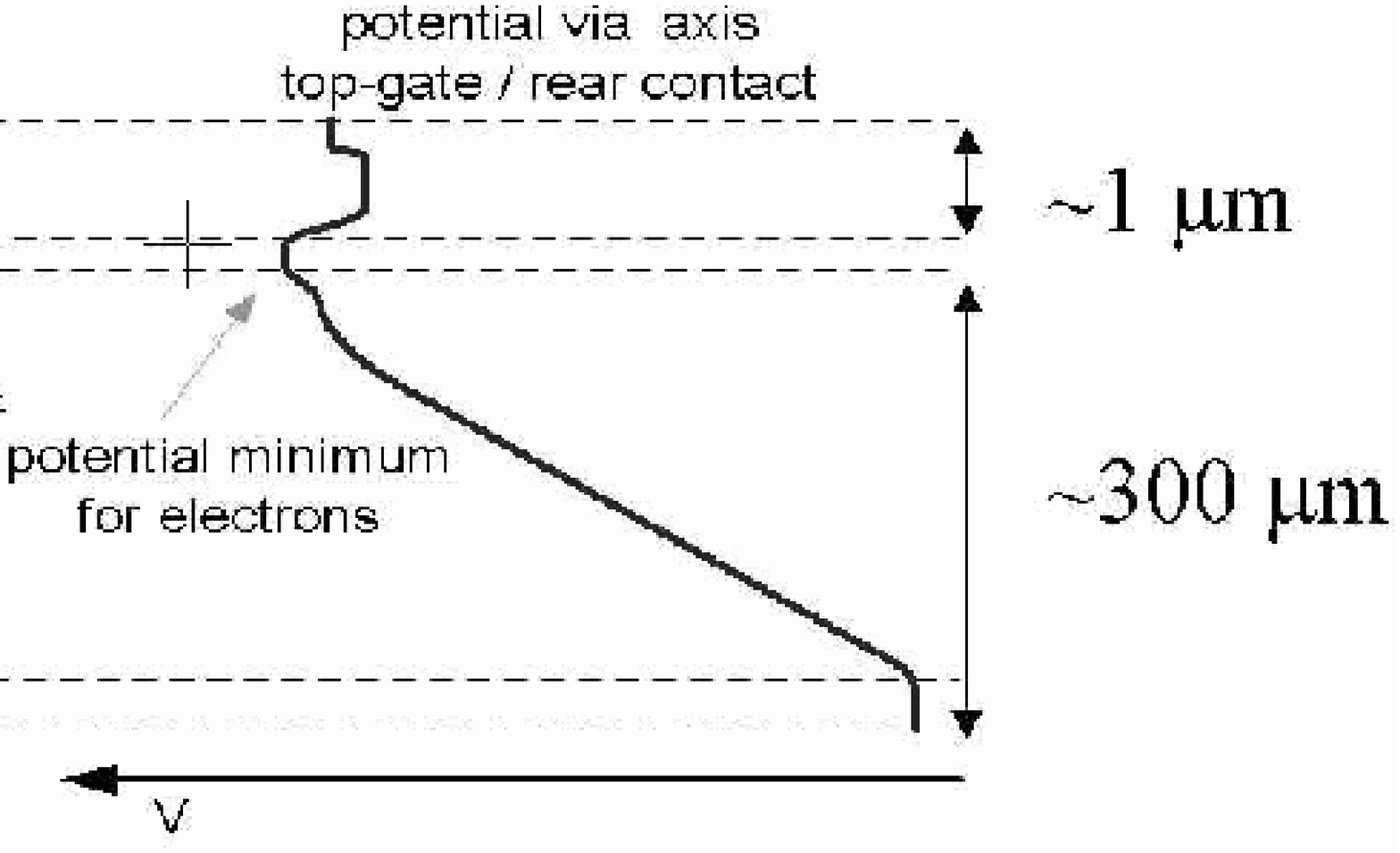}
\end{center}
\caption[]{\label{fig_depfetprinciple} Cross-section of a DEPFET
pixel (left side) and potential between top gate and rear contact
as function of depth (right side).}
\end{figure}

The DEPleted Field Effect Transistor structure~\cite{kemmer87},
abbreviated DEPFET, provides detection and amplification
properties jointly. The principle of operation is shown in fig.
\ref{fig_depfetprinciple}. A MOS or junction field effect
transistor is integrated onto a detector substrate. By means of
sidewards depletion~\cite{gatti84}, appropriate bulk, source and
drain potentials, and an additional deep-n-implantion a potential
minimum for electrons is created right underneath the transistor
channel ($\sim$ 1\,$\mu$m below the surface). This can be regarded
as an internal gate of the transistor. A particle entering the
detector creates electron-hole pairs in the fully depleted silicon
substrate. While the holes drift into the rear contact of the
detector, the electrons are collected in the internal gate where
they are stored. The signal charge leads to a change in the
potential of the internal gate, resulting in a modulation of the
channel current of the transistor.

The simultaneous detection and amplification feature makes DEPFET
pixel detectors very attractive for low noise
imaging~\cite{ulrici00,klein96}. For particle detection the use of
very thin ($\sim50\,\mu$m) detectors operated with very low power
consumption should be possible. The low noise operation is
obtained because the capacitance of the internal gate can be made
very small (several 10\,fF), much smaller than the cell area
suggests. Furthermore, no external connection circuitry to the
first amplification stage is needed. External amplification enters
only at the second level stage. This leads to an excellent noise
performance already at room temperature. The pixel delivers a
current signal which is roughly proportional to the number of
collected electrons in the internal gate. Signal electrons as well
as electrons accumulated from bulk leakage current must be removed
from the internal gate after readout. Clearing is obtained by
periodically applying a positive voltage pulse to a clear contact.
Other clear mechanisms have also been studied \cite{ulrici03}. The
question, whether the internal gate is completely emptied from
electrons upon a CLEAR pulse (complete CLEAR) is an important one,
both for very low noise operation in X-ray astronomy with XEUS and
for fast pedestal subtraction in a LC detector. With complete
clearing the statistical fluctuations in the number of electrons
in the internal gate as well as switching noise (kT/C noise) are
absent.

\begin{figure}[h]
\begin{center}
\includegraphics[width=0.23\textwidth]{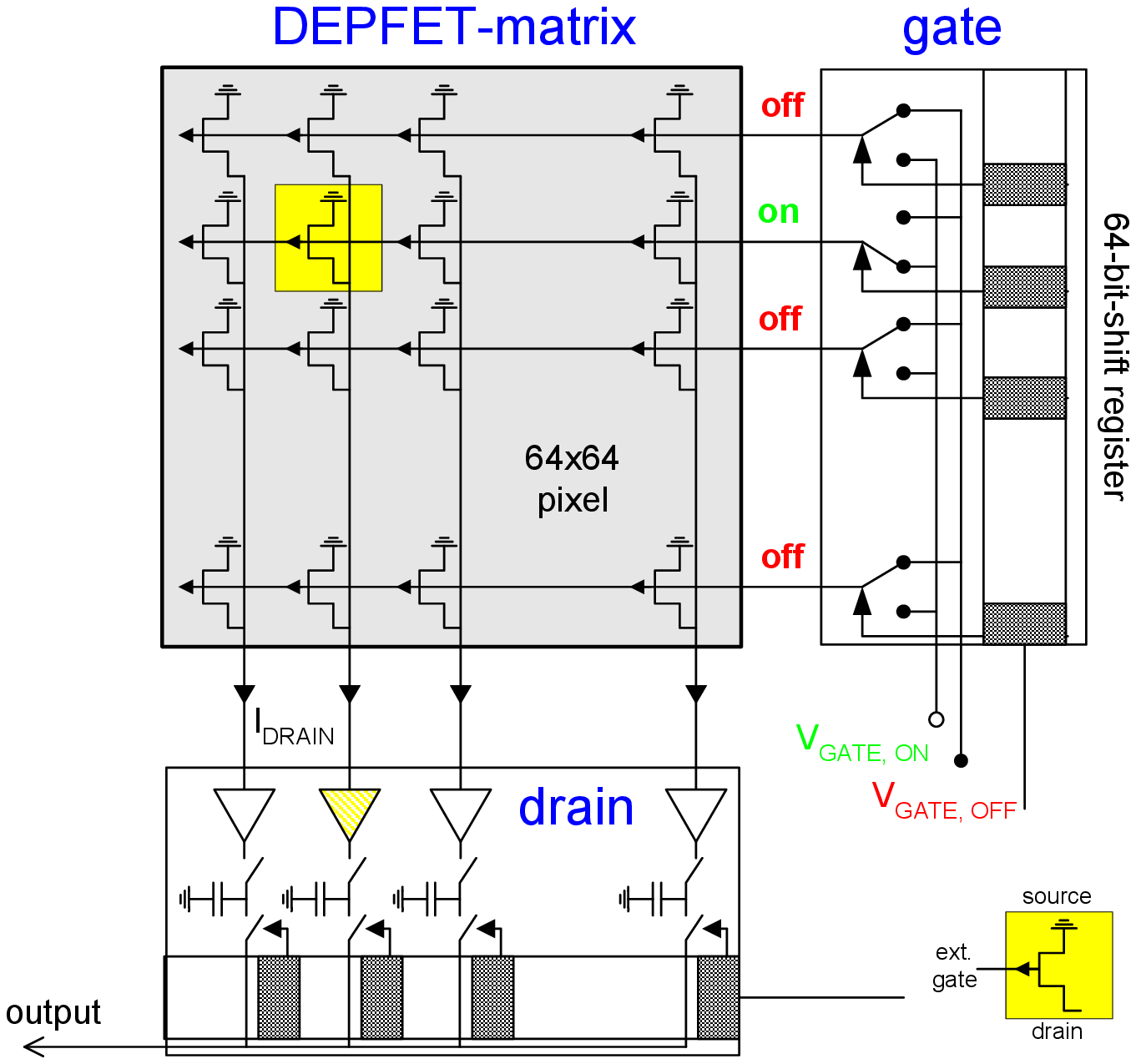}
\includegraphics[width=0.22\textwidth]{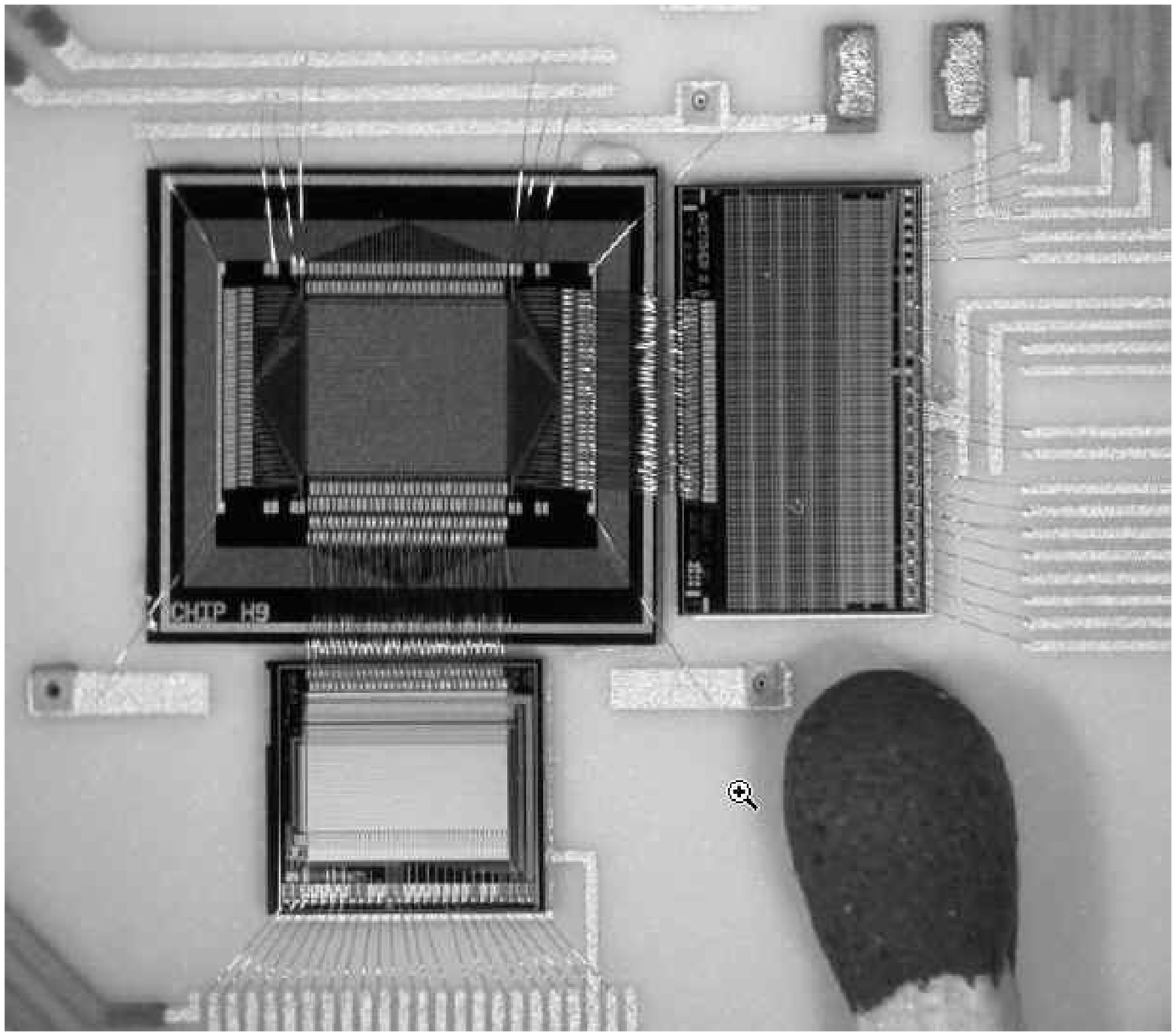}
\end{center}
\caption[]{\label{DEPFET_matrixoperation} Principle of operation
(left) and photograph (right) of a DEPFET pixel matrix showing the
steering IC for gate and clear control and readout IC containing
the current amplification stage at the bottom.}
\end{figure}

A DEPFET matrix is operated as shown in fig.
\ref{DEPFET_matrixoperation}. Rows are selected by applying a
voltage to the external gate of a row. Drains are connected
column-wise. The drain current of each pixel in a selected row is
detected and amplified in a dedicated amplification circuit.
Pedestals are taken at the beginning of an exposure cycle and
subtracted off-line. Finally, clear pulses are applied to the
clear contacts to empty the internal gates. On the right of fig.
\ref{DEPFET_matrixoperation} a photograph of a DEPFET-Matrix
hybrid assembly as used for imaging is shown. Both, gate-on/off
and CLEAR pulses are issued from the sequencer chip on the right.

\section{Performance of DEPFET single pixels and matrices}
\begin{figure}[h]
\begin{center}
{\includegraphics[width=0.5\textwidth]{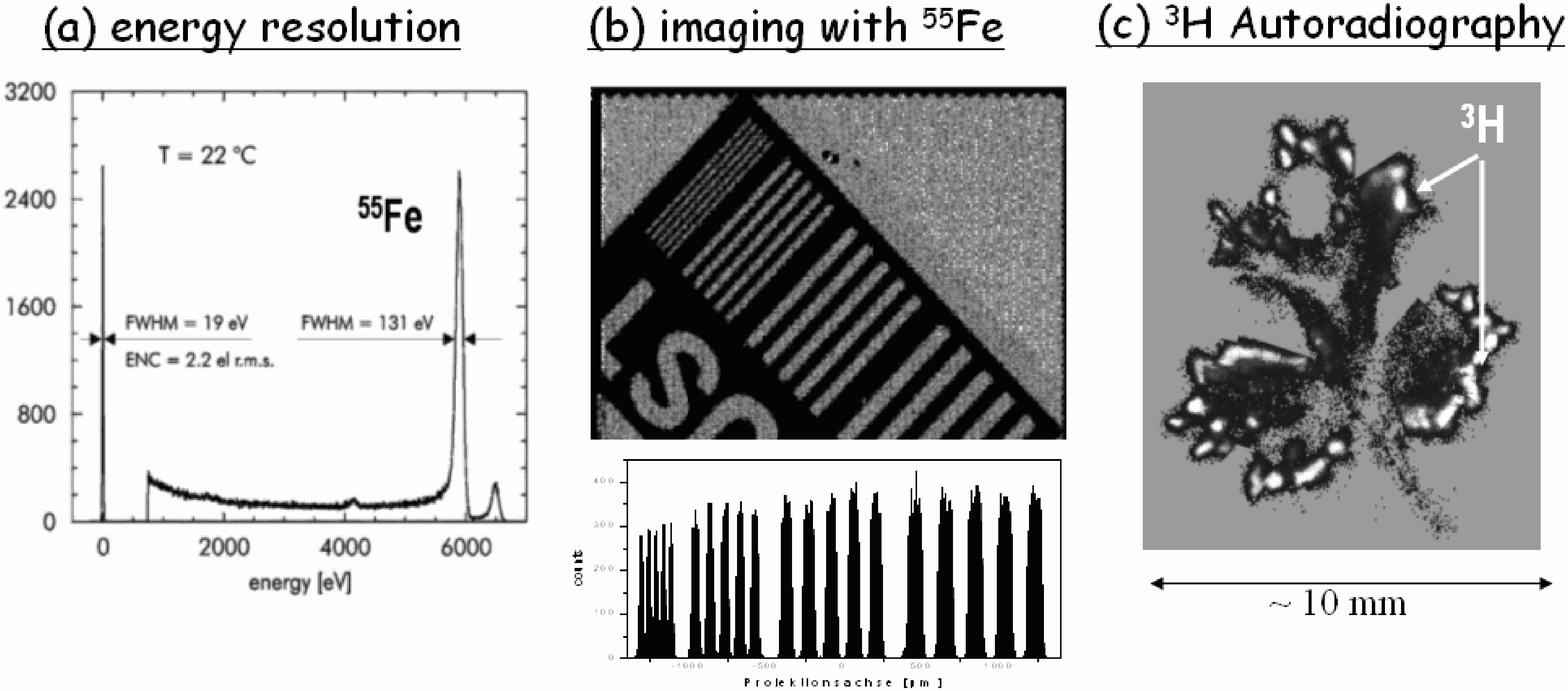}}
\end{center}
\caption[]{\label{DEPFET-Performance} (a) $^{55}$Fe energy
spectrum measured with a single pixel DEPFET structure, (b) image
of a tungsten test chart with line spacings down to 25$\mu$m, (c)
autoradiogram of a $^3$H-labelled leaf.}
\end{figure}

Figure \ref{DEPFET-Performance} summarizes the performance figures
obtained with DEPFET single pixels and with a 64x64 DEPFET matrix,
respectively. Fig. \ref{DEPFET-Performance}(a) shows the measured
energy spectrum obtained with a single DEPFET pixel structure at
room temperature. Using circular structures of a recent production
energy resolutions of 131 eV at 6 keV ($^{55}$Fe, K$_\alpha$ peak)
have been measured, originating from a Fano noise contribution of
14e and a DEPFET noise contribution of 2.2e. Figure
\ref{DEPFET-Performance}(b) displays the image of precision slits
in a tungsten test chart, the smallest of which are 25$\mu$m wide
at 50$\mu$m pitch. The projection of the image is shown below in
fig. \ref{DEPFET-Performance}(b). An evaluation of the measured
structure results in a width of the point spread function of
$\sigma$ = 6.7 $\pm$ 0.7 $\mu$m using an $^{55}$Fe (6 keV) X-ray
source and $\sigma$ = 4.3 $\pm$ 0.8 $\mu$m using a $^{109}$Cd (22
keV) source, corresponding to MTF$_{30\%}$ values of 37 lp/mm and
57 lp/mm, respectively \cite{ulrici03}. The corresponding width of
the point spread function for Tritium imaging derived from these
measurements is $\sim$7$\mu$m. Fig. \ref{DEPFET-Performance}(c)
finally shows the autoradiogram of a tritium-labelled leaf imaged
with a 64x64 DEPFET pixel matrix. The ability to simultaneously
distinguish different radio labels ($^3$H and $^{14}$C) in real
time has also been demonstrated \cite{ulrici03}.
Beyond the development for autoradiography, we are presently
developing DEPFET pixels for imaging with the X-ray telescope of
the planned XEUS satellite of the European Space Agency ESA
\cite{strueder03,XEUSref}, and for a particle micro vertex tracker
for a future Linear Collider like TESLA \cite{teslatdr,PRC03}.
While for XEUS very good energy resolution (ENC $<$ 4e) at relaxed
frame times (1.2 ms) is required, at TESLA frame times of 50$\mu$s
for 520x4000 pixels and 50$\mu$m thin detectors are needed.

\section{Developments for a Vertex detector for TESLA}
For a pixel detector at TESLA, DEPFET pixels can offer
\begin{itemize}
\item 50$\mu$m thin detectors with S/N values still in excess of
$\sim$50.

\item A total material budget (sensor, chips and frame) of
0.11$\%$X$_0$: The 50$\mu$m thin sensors sit in a perforated
300$\mu$m silicon frame (see section \ref{thinning}).

\item A cell size of 25x25$\mu$m$^2$ or similar.

\item A fast readout with 50 MHz line rate and 25 kHz frame rate
for 520x4000 pixels per module

\item Low power consumption of $<$5W for the entire detector with
5 barrel layers as proposed for TESLA \cite{teslatdr}.
\end{itemize}

\subsection{New DEPFET Sensors}
\begin{figure}[h]
\begin{center}
\includegraphics[width=.5\textwidth]{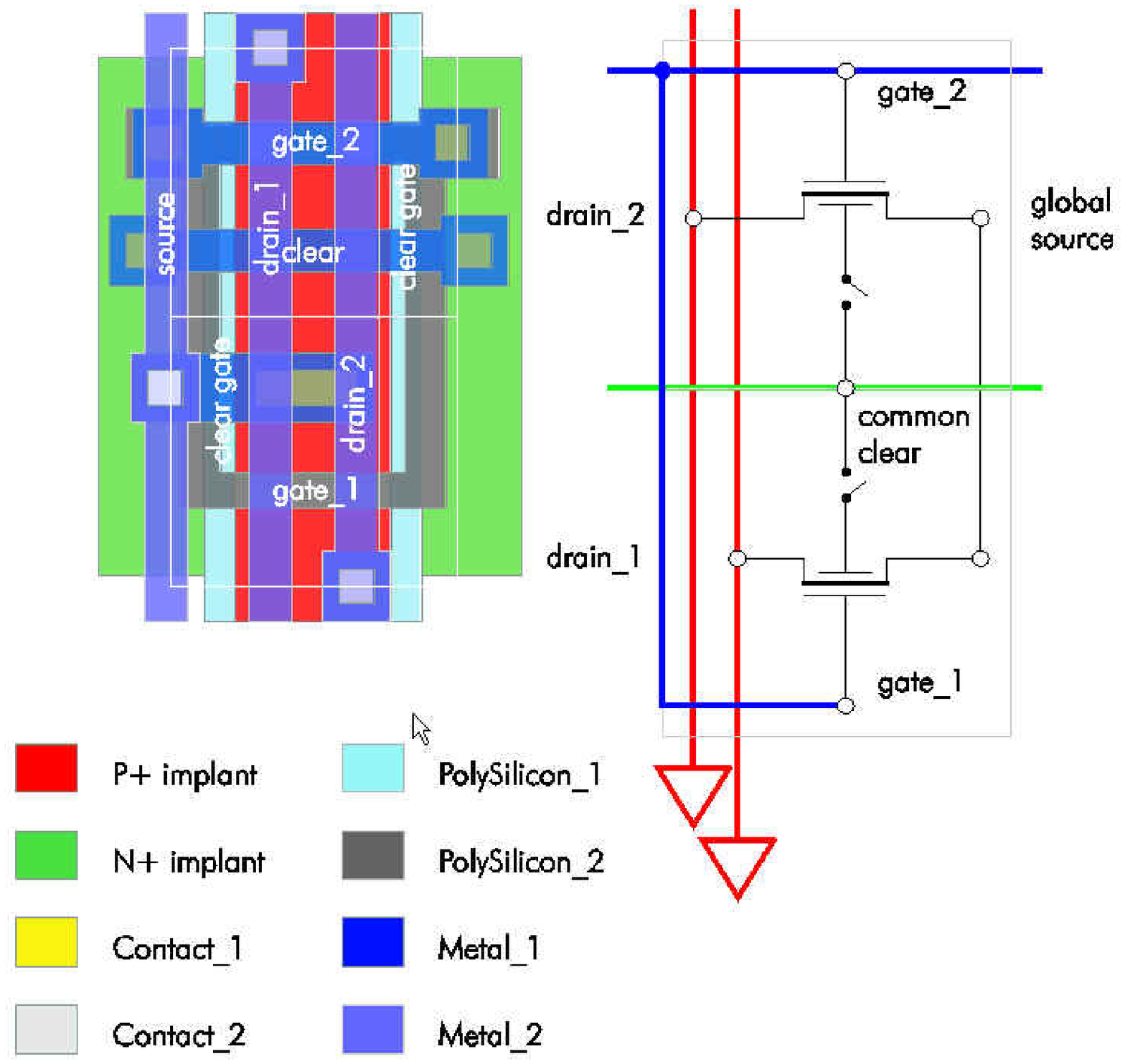}
\end{center}
\caption[]{\label{DEPMOS-Layout} Layout (left) and equivalent
circuitry diagram (right) for a linear pixel double-cell. Each
white rectangle surrounds a DEPMOS with CLEAR (N$^+$ green), a
common SOURCE (center P$^+$) and two DRAIN contacts (top and
bottom P$^+$). The double cell has a common source implantation.
The effective region for one cell is 20$\times$25$\mu$m.}
\end{figure}

Based upon the encouraging performance of DEPFET single structures
and large matrices, new DEPFET structures tailored to the needs of
the XEUS (75$\mu$m diameter circular structures) and TESLA (small
20$\times$25$\mu$m$^2$ linear structures) projects have been
fabricated at the MPI Semiconductor Laboratory in Munich using a
design with 2-metal layers. Among the main design goals were
\begin{itemize}
\item smaller pixels by a factor of 2 in both linear dimensions:
This enforced the change from closed circular JFET geometries,
used so far, to linear MOSFET structures, for which smaller
devices with better homogeneity and reproducibility over a large
area can be made with the existing technology. In order to reduce
the size further cells containing two pixels with a common clear
structure, a common source and two drain contacts have been
produced (see fig. \ref{DEPMOS-Layout}). One row addresses always
a pair of pixels with the drains connected to two separate column
like output busses. \item One of the main drawbacks of the
previous structures was the uncomplete clearing of the internal
gate, limiting the ability to obtain the achieved single pixel
noise figures below 5e also in large matrices. While complete
clearing of the internal gate is mandatory for low energy X-ray
imaging with XEUS, it is also necessary for operation with TESLA
speeds, because here signal and pedestal currents are on-chip
subtracted with a CLEAR pulse being applied between subsequent
signal and pedestal samplings (c.f. section \ref{chips}. In order
to facilitate this operation a ClearGate structure has been added
around the entire pixel cell, which is activated shortly ($\sim$
10 ns) after the CLEAR pulse. A ClearGate structure with the
required dimensions is very difficult to fabricate with a JFET
realization of the DEPFET transistor.
\end{itemize}
Simulations have confirmed that no potential pockets, which can
trap signal charges during charge collection and clearing, are
present in the design.

\begin{figure}[h]
\begin{center}
\includegraphics[width=0.4\textwidth]{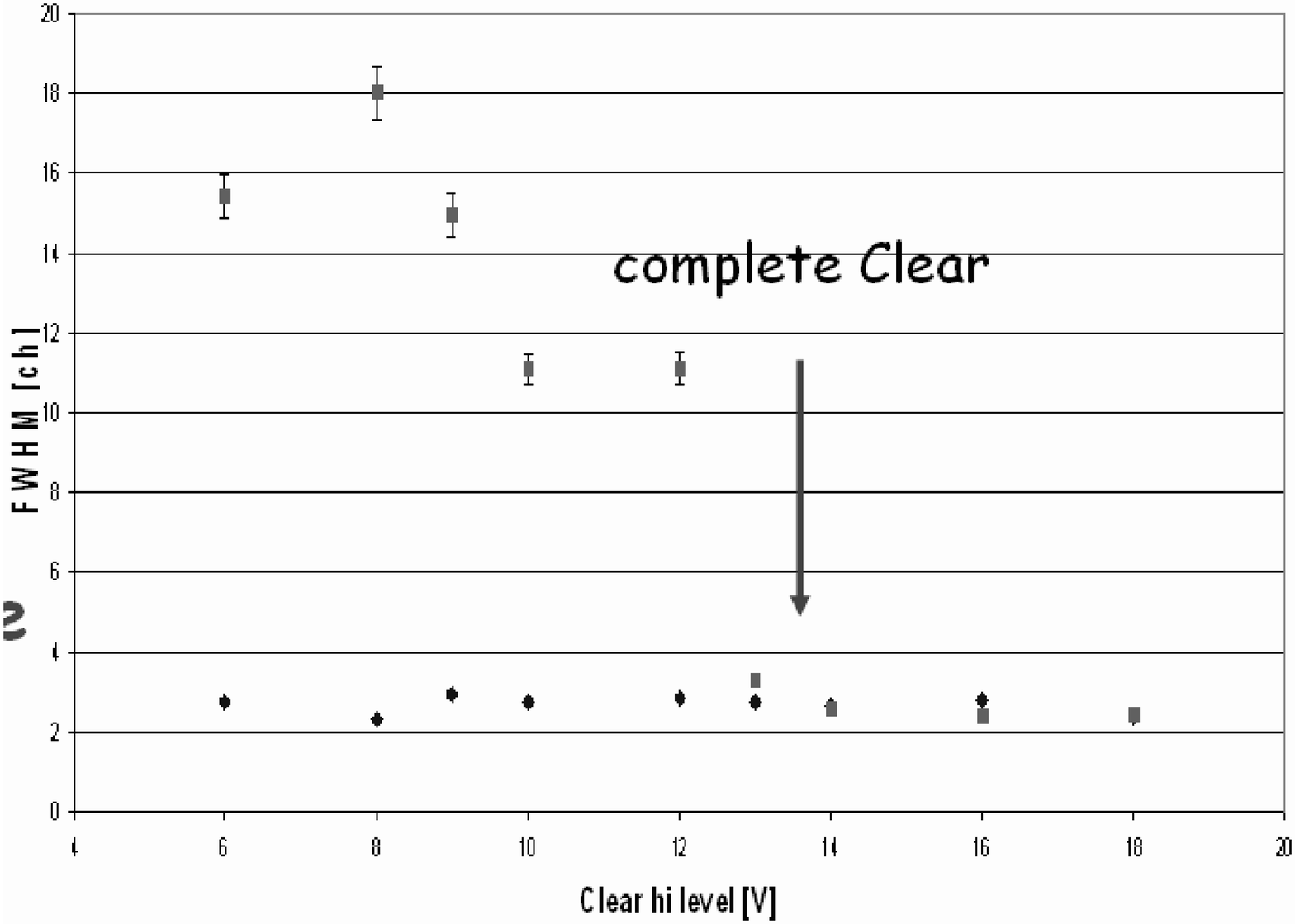}
\end{center}
\caption[]{ \label{complete-CLEAR} For Clear levels larger than
14V the width of the noise peak of different sampling sequences
(see text) become equal. This is a first evidence for a complete
clearing of the internal gate.}
\end{figure}

The smallest structures of the present production are about 20x30
$\mu$m$^2$. First measurements on single structures using an
$^{55}$Fe X-ray source have confirmed the excellent noise
performance of previous structures. With the low noise circular
structures designed for the XEUS project, an ENC of 2.2 e$^-$ and
131 eV FWHM for the K$_\alpha$ line have been measured at room
temperature. The charge to current amplification has been measured
to be $\sim$400pA/e$-$ on the first structures which have been
evaluated. Figure \ref{complete-CLEAR} indicates that complete
clearing has been achieved. Shown is the width of the noise peak
obtained by measuring the DC DEPFET drain current using two very
different R/O operations: (a) one CLEAR and then taking 500 R/O
samples, (b) taking 500 x (CLEAR + R/O) samples. If the clearing
is complete the widths of the noise peaks in both cases should be
the same. This is the case for sufficiently large CLEAR pulses
with sufficient duration time (e.g. $>$ 14V, $>$ 100 ns pulse
width).

Very first studies on DEPFET matrices indicate DEPFET current
dispersions of about 7.5$\%$ over a 16$\times$128 pixel matrix
operated without clearing. The space resolution has not yet been
measured in test beams yet. Extrapolating from measurements with
previous DEPFET pixel matrices with cell dimensions of
50$\mu$m$\times$50$\mu$m we expect space resolutions for minimum
ionizing particles with these new sensors in the order of
2-4$\mu$m, depending on the charge sharing between pixels.

\subsection{Wafer thinning technology}\label{thinning}

\begin{figure}[h]
\begin{center}
\includegraphics[width=0.5\textwidth]{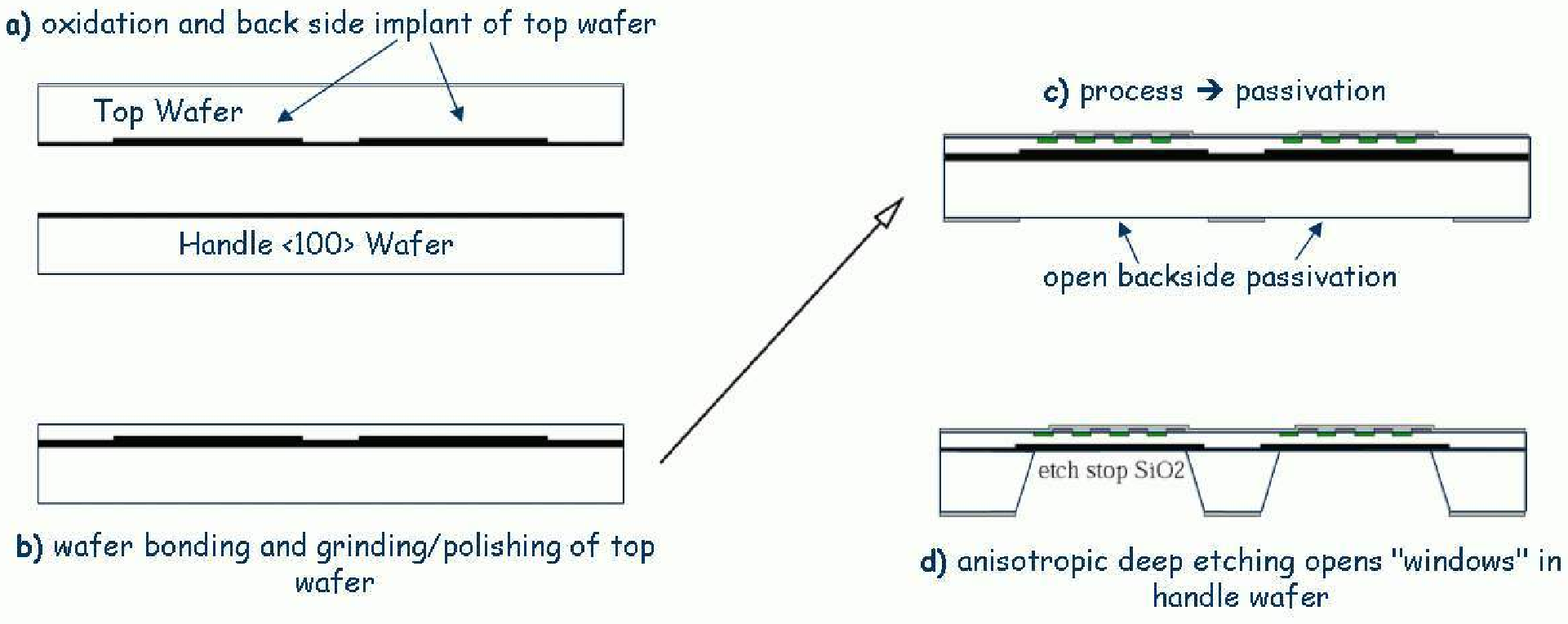}
\end{center}
\caption[]{ \label{thinning-process} Process sequence of wafer
thinning and DEPFET production: (a) the top (sensor) wafer
contains already processed diodes at the bottom, the bottom wafer
serves as a mechanical support, (b) after bonding, the top wafer
is ground and polished to 50$\mu$m thickness, (c) final processing
of DEPFET sensors on the top wafer, (d) etching of the backside
wafer which stops at the interface oxide between the two wafers}
\end{figure}

\begin{figure}[hbt]
\begin{center}
\includegraphics[width=.5\textwidth]{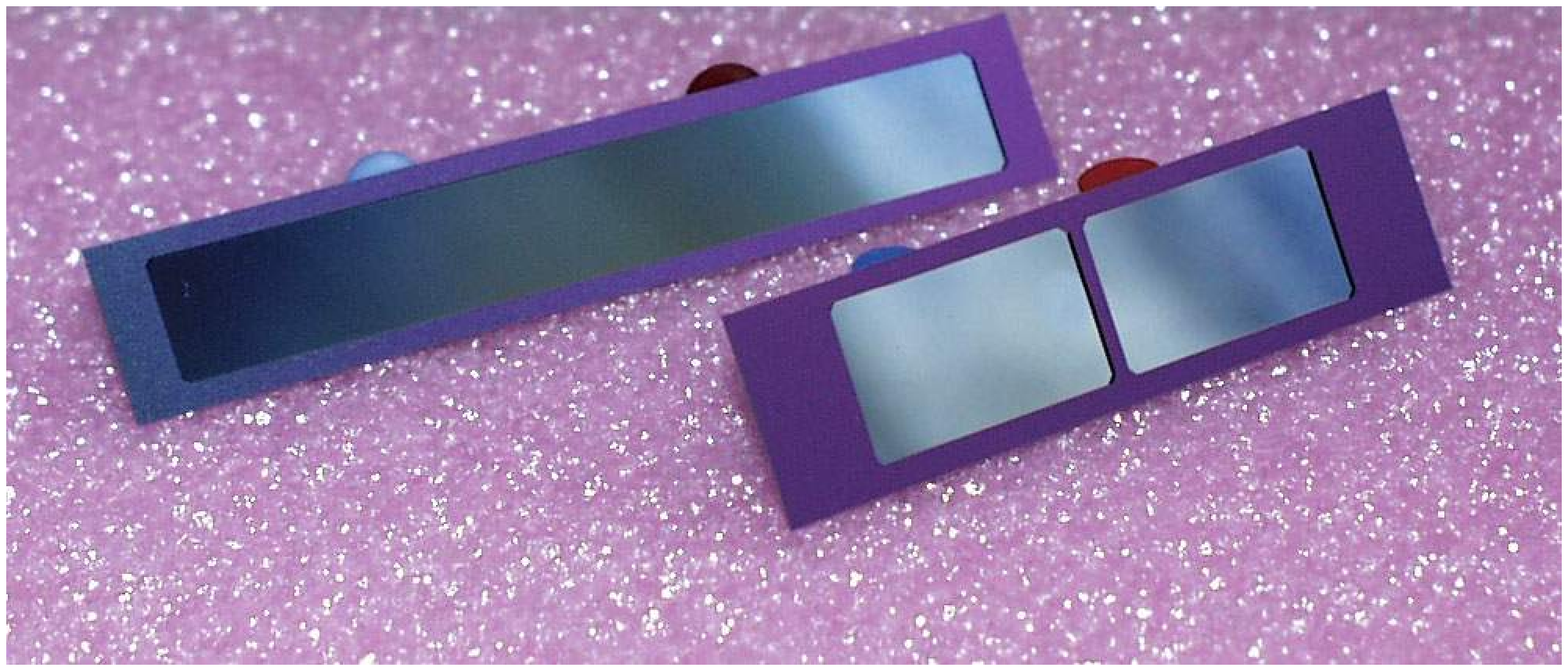}
\end{center}
\caption[]{ \label{thinned-silicon} Results of the thinning
technology development for mechanical samples. The size of the
upper part is 800x104mm². A 300\, $\mu$m thick silicon substrate
which is thinned down to a thickness of 50\,$\mu$m in the central
area.}
\end{figure}

Thinning of DEPFET wafers using anisotropic etching has been
developed successfully in collaboration with the MPI f\"ur
Mikrostrukturphysik in Halle/Germany \cite{laci03}. The processing
steps using bonding of a handle wafer are shown in fig.
\ref{thinning-process}. A thin 50$\mu$m structure which is
stiffened by a frame of thicker silicon ($\sim$ 350$\mu$m) is
obtained. The detector remains sensitive also in these thicker
frame regions. Active Si diodes have been thinned this way with
very satisfactory results and leakage currents less than 1
nA/cm$^2$ \cite{laci03} have been measured. A photograph of
thinned passive, module sized samples is presented in fig.
\ref{thinned-silicon}.

\subsection{Readout chips for DEPFET modules at a LC}\label{chips}
The readout principle for a DEPFET matrix of fig.
\ref{DEPFET_matrixoperation} is unchanged for a much faster
operation at a Linear Collider. However, a module of 520x4000
pixels is read out column-wise to both top and bottom sides at a
continuous line rate of up to 50 MHz (40$\mu$s frame time) in the
innermost layer, taking 25 frames per train crossing time ($\sim$
1 ms). Both, the row addressing sequencer chip SWITCHER for
(external) gate on/off and (internal gate) clear and the
column-based current readout chip (CURO) for amplification and
sparsified current readout must comply to this rate. During one
line cycle (20 ns), sampling of signal+pedestal, CLEAR, and
sampling of pedestal is required.

\begin{figure}[h]
\begin{center}
\includegraphics[width=.45\textwidth]{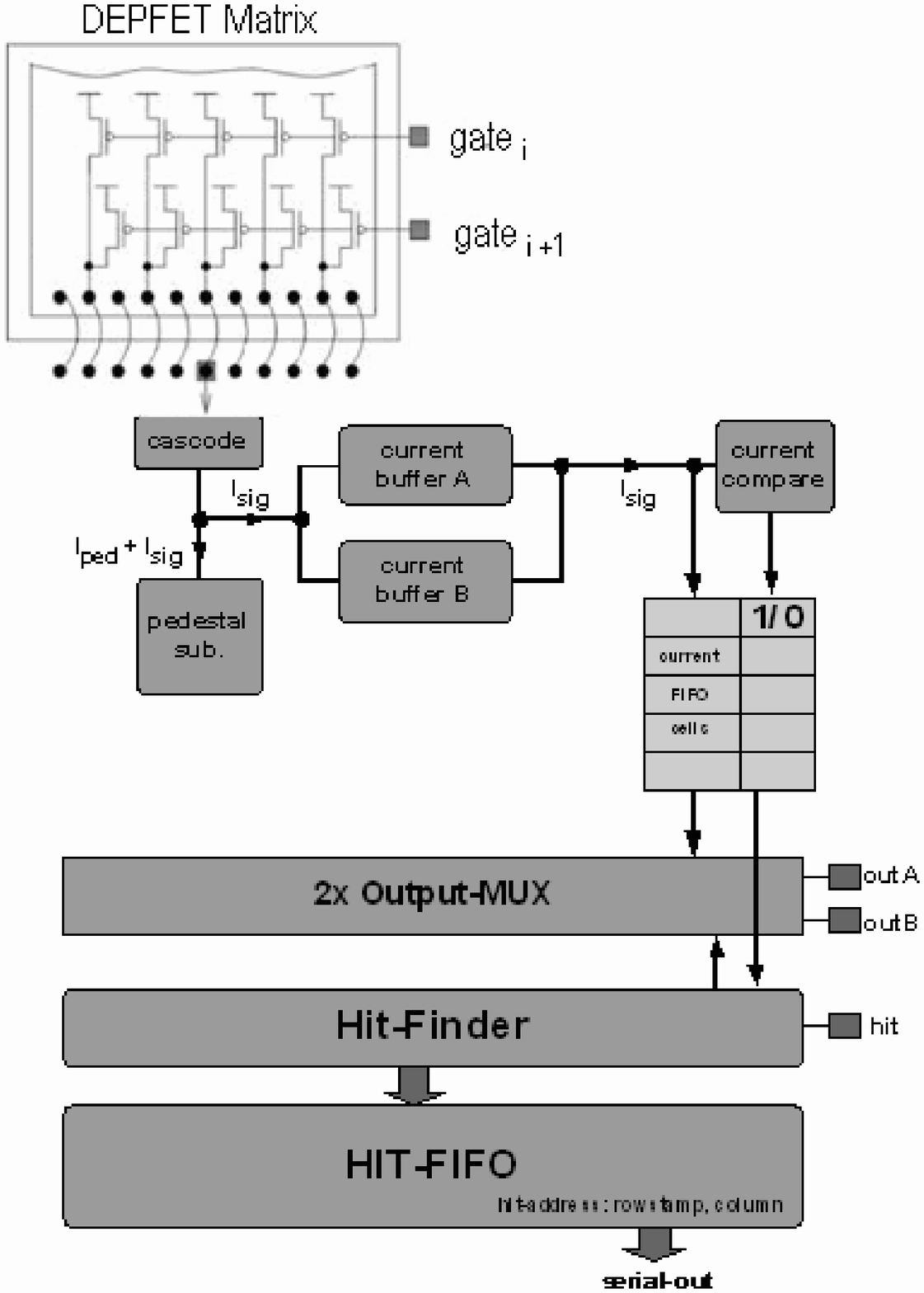}
\end{center}
\caption[]{ \label{TESLA-readout} Readout scheme for DEPFET pixel
matrices at a LC.}
\end{figure}

The block diagram of the readout chip CURO is shown in fig.
\ref{TESLA-readout}. A cascode stage at the input of the readout
chip keeps the drain lines at a constant potential, such that the
DEPFET current does not need to charge the relatively large bus
capacitance. A row is selected for readout after the charge
accumulation interval. The selected DEPFET transistors output a
signal current superimposed to a pedestal current. This current
$I_{ped}+I_{sig}$ is stored in a current memory cell for every
column. After a complete CLEAR of the row, and assuming negligible
leakage currents during the time between two sample cycles (10
ns),
the pedestal current $I_{ped}$ is measured by a second sampling
cycle in the selected row. The pedestal subtracted signal is then
obtained by summing the stored (negative) signal+pedestal current
with the pedestal current. The signal current $-I_{sig}$ is stored
in an analog FIFO of several current memory cells. A fast current
comparison with a programmable hit threshold generates a digital
hit pattern which is simultaneously recorded in a digital FIFO.
Analog FIFO cells with no hits can be switched off to save power.
While the analog FIFOs are filled with events containing at least
one hit, a fast scanner searches for hits in the digital FIFO. The
corresponding analog values are selected with a multiplexer and
fed to one or several ADCs.

\begin{figure}[h]
\begin{center}
\includegraphics[width=.25\textwidth]{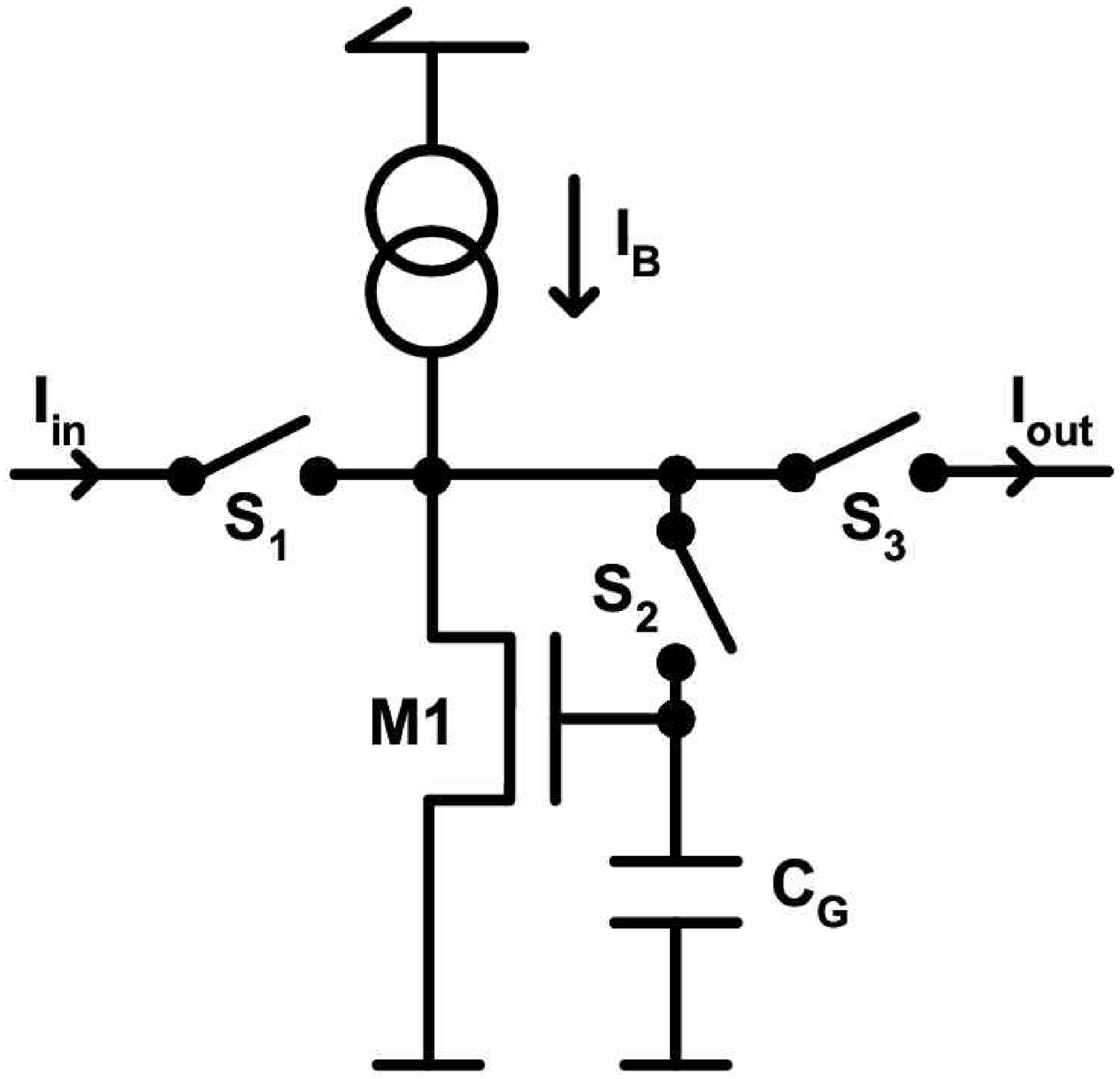}
\end{center}
\caption[]{\label{memory-cell} Principle of the current memory
cell. $C_{G}$ is the gate capacitance of the transistor M1.}
\end{figure}

Crucial elements of this design are the current memory cell, the
hit scanner, and the current comparator. They have all been tested
using a test chip fabrication. The current memory cell (fig.
\ref{memory-cell}) is adopted using the switched current technique
\cite{SWITCHEDCURRENT}. A current is stored in three phases:

\begin{enumerate}
\item{{\em Storage Phase:} S1 and S2 are closed, S3 open. The gate
capacitance of the transistor M1 is charged until the device
provides the combined input and bias current ($I_{M1}=I_{in} +
I_B$).} \item{{\em Sampling Phase:} S2 is opened. The gate voltage
and therefore the transistor current ideally remain unchanged.}
\item{{\em Transfer Phase:} Immediately after sampling S1 is
opened and S3 closed. As the current through M1 is still
$I_{M1}=I_{in} + I_B$, $I_{in}$ must be delivered by the output
node.}
\end{enumerate}

Thus, in the ideal case $I_{out}= - I_{in}$. However, this simple
circuit suffers from several non-ideal effects like
charge-injection of the sampling switch S2 and the limited output
conductance of the transistor M1 and the biasing current source.
Therefore in the real case $I_{out} = - I_{in} + \delta I$ where
$\delta I$ indicates the error made by the sampling process. Many
techniques to cope with these deficiencies have been treated in
the literature \cite{ACCURATE_CITE}. Here, cascode techniques have
been used for the sampling transistor and the current source to
decrease the output conductance. The implemented circuitry uses a
two stage design to cancel charge injection and to achieve a high
dynamic range with a small storage error. Measurements using a
test chip at a 25MHz sampling rate (limited by the test setup)
show a differential non-linearity of $0.1\%$ over a sufficiently
large dynamic range of 10 $\mu$A corresponding to several mips. At
the cost of linearity the total dynamic range can be as large as
100$\mu$A, corresponding to the charge of 20 mips or more,
depending on the g$_q$ of the DEPFET transistor. The total
standing current in the memory cell operated at a supply voltage
of $2.5$V is $150\mu A$ so that for the input stage the calculated
power consumption is $2mW$ per DEPFET column. The {\em hit finder}
uses a binary tree structure as first proposed in \cite{mephisto}
to find two hits out of a pattern of 128 digital inputs within one
clock cycle. Measurements on the hit finder circuitry implemented
on the test chip confirm that a speed of 50MHz is easily
achievable in the chosen technology. A full chip for a 64x128
DEPFET matrix has been submitted using radiation tolerant design
rules. One important of the many advantages of the switched
current based technique is its little dependence on smaller and
smaller becoming operating voltages in advanced IC technologies.

\begin{figure}[h]
\begin{center}
\includegraphics[width=.5\textwidth]{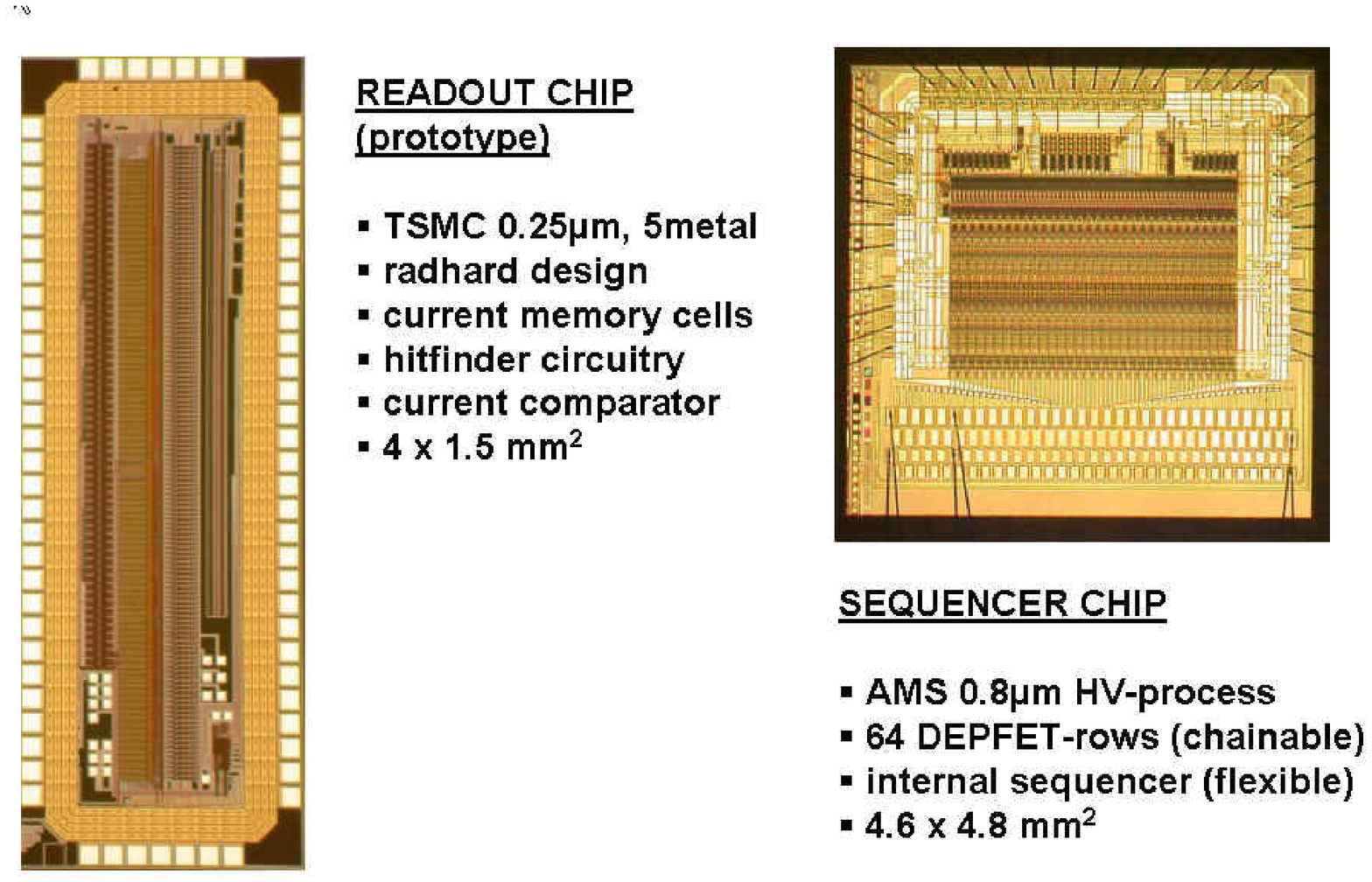}
\end{center}
\caption[]{ \label{prototype-chips} (a) Readout chip prototype
containing the main building blocks of a readout chip for TESLA
(current memory, hit finder, comparator), and (b) sequencer chip
for TESLA and XEUS.}
\end{figure}

The sequencer chip (SWITCHER) generates high voltage ($<$20V)
signals for gate on/off and CLEAR of rows. In order to provide a
wide voltage range for testing of the first matrices, a special
'high voltage' 0.8$\mu m$ technology has been chosen. The chip
(photo in fig.~\ref{chips}b) contains 64 channels with analog
multiplexers, a digital sequencer and control logic. It is
designed to be suited for of all fabricated DEPFET matrices for
the different applications including 50MHz operation at TESLA.
Several chips can be daisy chained to control larger arrays.
Measurements on the fabricated prototype chip show that it is
fully functional at 30 MHz speed. By addressing 2 pixels at the
same time (see fig. \ref{DEPMOS-Layout}) the target line rate for
TESLA would be 25 MHz, i.e. a time of 40 ns per line, during which
sampling of the signal current, CLEAR, and sampling of the
pedestal current must be done.

The total power consumption has been calculated using measured
currents and voltages of prototype detectors and chips, and
assuming that rows can be switched off while not being read out.
Scaled to the total pixel vertex detector \cite{teslatdr} of TESLA
with 5 layers, the sensor
is expected to have a power consumption of 0.3W, while the ICs
consume $\sim$3-4W (SWITCHER) and $\sim$1-2W (CURO), respectively.
Hence, we consider a total power budget of less than 5W for the
entire detector to be feasible. This figure renders air cooling or
alternative low mass cooling to be possible, hence offering a very
low total material budget ($\ll$ 1$\%$ X$_0$).

\section{Conclusion}
In contrast to other trends in monolithic or semi-monolithic pixel
detector developments (see e.g. \cite{wermes03,deptuch03}),
partially using incomplete charge collection, DEPFET pixels excel
in excellent energy (ENC = 2.2e, $\sigma_E$=131 eV) and space (4.3
$\mu$m at 20 keV X-rays) resolutions. This can be exploited in a
variety of applications. In biomedical autoradiography the
excellent imaging performance of DEPFET pixels has been
demonstrated by space resolved imaging of Tritium-labelled
biological tissue. A spatial resolution for $^3$H of $\sim$7$\mu$m
as inferred from measurements with X-ray sources has been
concluded. For the future use of DEPFET pixels for X-ray astronomy
within the ESA-XEUS mission or for high energy particle vertex
tracking, smaller and, in terms of resolution, still better
performing DEPFET structures have been developed. For use at a
future Linear Collider a R/O architecture integrated circuits have
been developed with speeds larger by a factor 1000 than achieved
so far. These include a current based R/O chip with fast and low
noise current memory cells and a current-hit finder.

\section*{Acknowledgment}
The authors would like to thank the technology crew of PN sensor
GmbH and MPI at the Semiconductor Laboratory.



\end{document}